# Decoration of exfoliated black phosphorus with nickel nanoparticles and application in catalysis

Maria Caporali,[a] Manuel Serrano-Ruiz,[a] Francesca Telesio,[b] Stefan Heun,[b] Giuseppe Nicotra,[c] Corrado Spinella[c] and Maurizio Peruzzini[a]

**Nickel nanoparticles were dispersed on the surface of exfoliated black phosphorus and the resulting nanohybrid Ni/2DBP showed an improved stability respect to pristine 2D BP when kept in ambient conditions in the darkness. Ni/2DBP was applied as catalyst in the semihydrogenation of phenylacetylene and exhibited high conversion and selectivity towards styrene. These features were preserved after recycling tests revealing the high stability of the nanohybrid.**

Black phosphorus (BP) has a layered structure analogue to graphite, and lately it has been shown that, following a similar procedure to bulk graphite, it can be exfoliated down to the monolayer. In this way, a new member of the growing family of 2D materials, named phosphorene,[1,2] the all P-counterpart of graphene, was isolated. Single and few-layer BP (2D BP) have been obtained by either micromechanical cleavage (Scotch tape method)[3] or liquid exfoliation.[4] In our labs, good quality phosphorene flakes were prepared by sonicating BP microcrystals in dimethylsulfoxide (DMSO) keeping an inert atmosphere and in the dark,[5] see Figures S1-S4 for the STEM and EELS characterization.

Since its discovery, the new P-based 2D material attracted an enormous interest, as it is endowed with a direct and tunable band gap ranging from 0.3 eV (bulk) to 2.0 eV (monolayer) and a high carrier mobility[6] (up to 6000 $cm^2V^{-1}s^{-1}$), which prompted its application towards the production of nano-electronic devices, as for instance field effect transistors (FETs).[7] In spite of this huge interest, only an handful of reports have dealt with the surface functionalization of 2D BP and its application in chemical processes so far.[8] Based on first-principles calculations, a systematic study on the binding energy, geometry, magnetic moment, and electronic structure of several metal ad-atoms adsorbed on phosphorene has been carried out,[9] predicting that the immobilization of transition metals on the surface of phosphorene is feasible, while preserving its structural integrity. On these basis, we added dropwise a colloidal solution of nickel nanoparticles[10] to a suspension of 2D BP under vigorous stirring. The reaction was monitored by UV-Vis spectrophotometry. Spectra were acquired first on colloidal Ni NPs and on a suspension of 2D BP, giving two distinct behaviours, see Figure 1. Running UV-Vis right after mixing the two suspensions, the dispersion curve characteristic of colloidal Ni NPs disappeared, suggesting that Ni NPs were immobilized on the surface of BP.

[a.] CNR-ICCOM, Via Madonna del Piano 10, 50019 Sesto Fiorentino, Italy, maria.caporali@iccom.cnr.it; maurizio.peruzzini@iccom.cnr.it.
[b.] NEST, Istituto Nanoscienze-CNR and Scuola Normale Superiore, Piazza S. Silvestro 12, 56127 Pisa, Italy.
[c.] CNR-IMM, Istituto per la Microelettronica e Microsistemi, Strada VIII, 5, 95121 Catania, Italy.

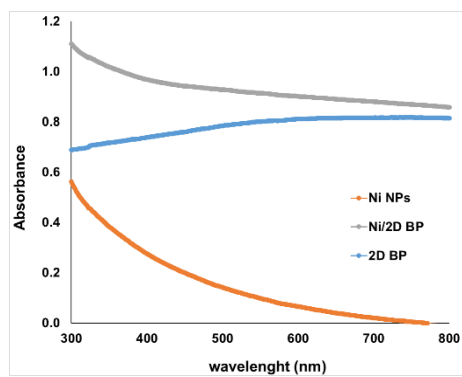

*Fig. 1* UV-Vis: a) 2D BP (light blue); b) Ni NPs (orange); c) Ni/2D BP (grey).

TEM analysis showed well dispersed nickel nanoparticles on the surface of black phosphorus nanosheets, see Figures 2 and S8. After immobilization, Ni nanoparticles preserved their size, average diameter of 11.9 ± 0.8 nm, and no aggregation took place, see for comparison Figure S5 for TEM image and size distribution of bare colloidal Ni NPs.

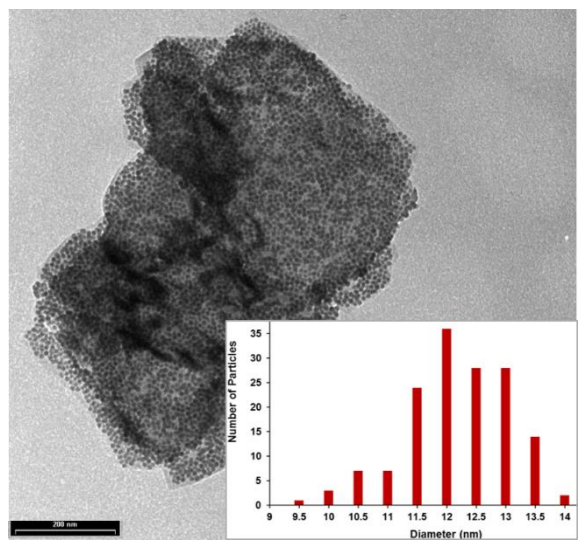

*Fig. 2* TEM image of Ni NPs supported on few-layer black phosphorus, P:Ni molar ratio 3:1. Scale bar: 200 nm. Inset: size distribution histogram of Ni NPs.

EDX performed on different points of the same sample confirmed the chemical identity of the 2D material and corroborated the presence of Ni and elemental P without any impurities see Figure S9. Afterwards, high angle annular dark field (HAADF) STEM was carried out to study the sample at the atomic level, see Figure 3.

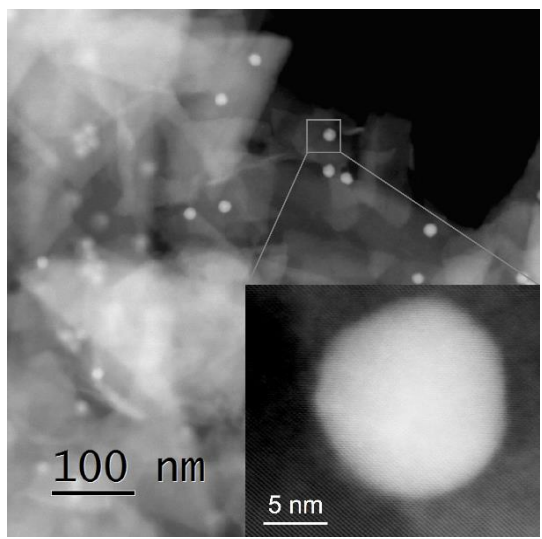

*Fig. 3* HAADF-STEM of Ni/2D BP with inset showing atomic resolution for 2D BP.

By the combination of both STEM and EELS analysis technique named spectrum imaging (SI),[11] we can disclose the chemical composition of the surface of the new nanocomposite. For this purpose, we acquired the experimental Ni $L_{2,3}$-edge energy loss near edge structure (ELNES), for Ni NPs and NiO NPs,[12] the latter taken as a reference sample, grey and red spectra in Figure 4, respectively, and compared with reference spectra from literature, blue and purple, respectively.[12] Our EELS spectra shown in Figure 4 were extracted from the SI dataset after background subtraction and multiple scattering deconvolution, in order to avoid artifacts due to specimen thickness. Here, besides the $L_3$ and $L_2$ peaks at 855 eV and 872 eV, respectively, a secondary peak is observed (black arrow I on grey and blue spectra). This secondary peak is present only in pure metallic Ni[13] and arises from a 4p-4d hybridized band[12] that is absent in nickel oxide (red spectrum). These results, compared with the spectrum acquired on NiO NPs, showing a typical bump around 869 eV[12] (see black arrow II on red and purple spectra of Figure 4), provide a strong confirmatory evidence that the prepared Ni NPs are not oxidized and add further evidence that the BP flakes are strictly decorated with not-oxidized metallic nickel nanoparticles. In keeping with this finding, the observed catalytic activity of the hybrid system, *vide infra*, confirms that no oxidation of Ni NPs has occurred.

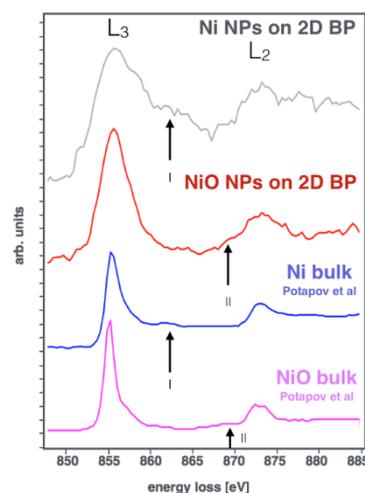

*Fig. 4* Ni $L_{2,3}$-edge Energy Loss Near Edge Structure (ELNES) for Ni/2D BP and NiO/2D BP, and the equivalent signals from bulk structures of Ni and NiO.[12]

The new nanohybrid was further characterized by Raman spectroscopy and Figure 5 compares the Raman spectrum of a freshly prepared sample of 2D BP (blue curve) with the same batch of 2D BP after functionalization with nickel nanoparticles (red curve).

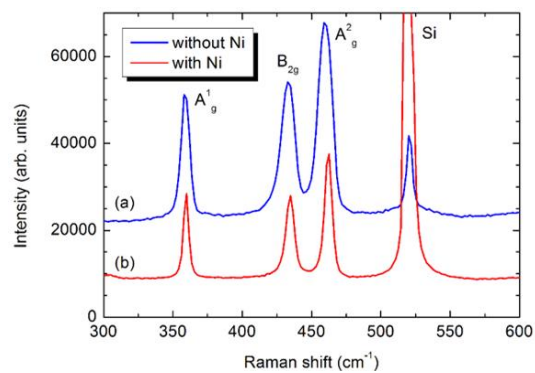

*Fig. 5* Comparison of the Raman spectra between a) pristine few-layer BP (blue curve); b) Ni/2D BP (red curve). The strong peak at 520 cm$^{-1}$ is due to the silicon substrate.

The spectrum of the nickel functionalized material reveals the three typical Raman bands of phosphorene at 360, 435 and 460 cm$^{-1}$, attributed to the $A^1_g$, $B_{2g}$, and $A^2_g$ vibrational modes respectively,[14] confirming the orthorhombic crystalline structure of 2D BP. Since nickel is silent in Raman, the crystalline nature of the nanoparticles was confirmed by SAED and powder XRD, see Figures S6-S7, which showed that the nanoparticles have the characteristic cubic face centred structure of crystalline nickel.

On the same sample AFM was measured. Looking at smaller structures, which are likely to be single flakes or few flake-aggregates, see Figure S11, we found their typical size to be a few hundred nanometers, and their thickness up to 50 nm. A typical flake is shown in Figure 6 (a), and two cross-sections are shown in Figure 6 (b) and (c). Knowing that a single Ni nanoparticle has a diameter around 12 nm, and using the same approximation as in Hersam's paper[8c] we can roughly estimate BP flake thickness. The object of Figure 6 (a), with a thickness from 14 nm to 32 nm, corresponds to a BP flake thickness of 2 to 20 nm, and being the BP layer spacing of 0.53 nm, the number of layers ranges from 4 to 38.

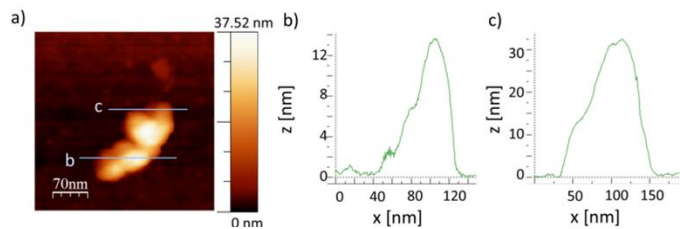

**Fig. 6** *(a) AFM image of a typical flake and b) and c) profiles along the two lines displayed in a).*

Powder X-Ray diffraction of Ni/2D BP (see Figure S10) shows the characteristic peaks of black P and additionally a broad peak at 2θ equal to 44.4° revealing the presence of nickel NPs. The other peaks at 51.8° and 76.4° also characteristic of nickel, are too weak to be identified. Thus, the surface functionalization with Ni NPs did not affect the structure and the crystallinity of 2D BP.

The environmental instability of exfoliated BP has been an issue so far, since in ambient conditions BP degrades completely in less than a week, preventing its applications in several fields.[15] Numerous strategies have been applied for the ambient protection of 2D BP, as the coating with $Al_2O_3$,[16] $MoS_2$[17] and recently with ionic liquids.[18] We were interested to study the influence, if any, of the functionalization with Ni NPs. A fresh suspension of Ni/2D BP was drop-casted on a TEM grid and on a $SiO_2$/Si support and exposed in the darkness to ambient oxygen and moisture. TEM and Raman measurements were acquired regularly over one week and in parallel, a sample of pristine 2D BP was examined as a reference. Surprisingly, while the latter was quickly degraded to molecular compounds, see Figures S12, the morphology of Ni/2D BP flakes was unaltered after one week as shown by TEM in Figure 7. Raman spectrum clearly displayed the typical peaks of BP.

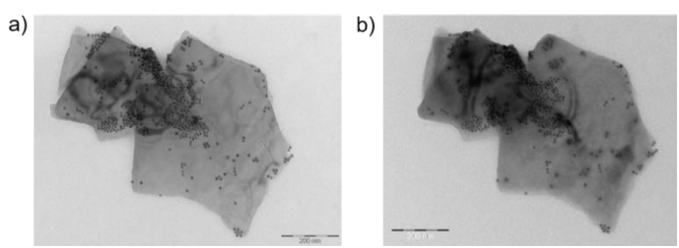

**Fig. 7** *Bright field TEM images of Ni/2D BP flake observed at a) time = 0, b) after one week. Scale bar: 200 nm.*

Being confident of the increased stability and aware that 2D BP possesses a higher surface to volume ratio than other 2D materials due to its puckered lattice configuration,[19] we investigated the catalytic activity of Ni/2D BP in a model reaction as the semihydrogenation of phenylacetylene to styrene. This is an important industrial reaction, because phenylacetylene is a poisoning impurity in styrene feedstocks that causes deactivation of the styrene polymerization catalyst.

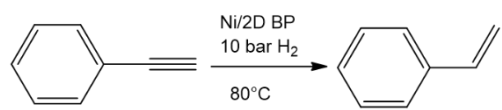

**Scheme 1** *Semihydrogenation of phenylacetylene to styrene.*

The conversion was quantitative with high selectivity towards styrene (see Table S1), outperforming known catalytic systems based on supported nickel nanoparticles.[20] A catalytic test with bare colloidal nanoparticles was run, see Table S1, and interestingly there was a drop from 92.8% to 78.6% in selectivity to styrene versus the test carried out with Ni/2D BP. Pristine few-layer BP was also tested and, as expected, did not show any catalytic activity.

The catalyst Ni/2D BP could be easily separated by centrifugation and reusability in the same process was tested. The conversion was quantitative till to the fourth run, and the selectivity remains unaltered, as shown in Figure 8. TEM inspection of Ni/2D BP recovered after the catalytic tests confirmed the presence of Ni NPs on the surface of the nanosheets, with an averaged dimension of 11.5 ± 0.9 nm, as before starting the catalysis, the BP flakes looked unchanged as well, see Figure S13. In comparison, unsupported Ni NPs lose their activity just after the first cycle.[21] These results indicate a strong metal-support interaction that infers high stability to nickel nanoparticles thus preventing their aggregation and leaching during the consecutive catalytic runs.

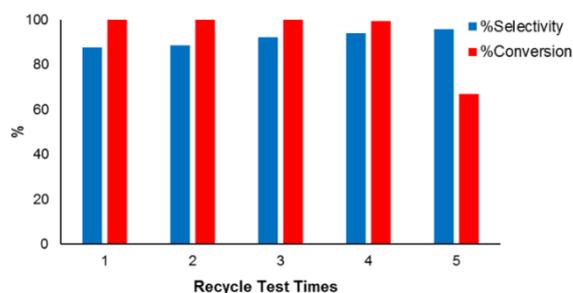

**Fig. 8** *Recycling tests. Reaction conditions: 10 bar $H_2$, 80°C, 2 hours, Ni NPs = 1.2 mol%, phenylacetylene = 0.3 mmol, 1.41 mg catalyst (15.2 wt% Ni). Both conversion and selectivity were evaluated by GC and GC-MS.*

Leaching tests were carried out to verify the presence in solution of both nickel and molecular compounds containing phosphorus, which might be possibly formed from the degradation of 2D BP during the process. Analysis by inductively coupled plasma mass spectrometer (ICP-MS) of the reaction mixture after catalyst separation showed that both Ni and P were absent at the ppm level, meaning Ni NPs are strongly immobilized on the surface of 2D BP which consequently acts as a suitable support and does not degrade to phosphorus oxyacids under catalytic conditions. Therefore, the observed decrease of the catalytic efficiency in the fifth run (67%) might be reasonably ascribed to nickel oxidation during work-up.

In summary, a new surface functionalization of exfoliated BP has been carried out with Ni NPs. Morphological analysis of the nanohybrid by AFM and electron microscopy was carried out and reveals the nanoparticles are homogeneously dispersed on the flakes' surface. Raman and XRD confirmed the nature and crystallinity of the material. A detailed study of the environmental stability of the new 2D material was carried out by means of TEM and Raman and resulted in an enhanced stability in ambient conditions towards the degradation to P-oxides and acids. Taking advantage of this property, Ni/2D BP was tested as reusable catalyst in the semihydrogenation of phenylacetylene, where it worked successfully exhibiting good conversion and selectivity to styrene in successive cycles, which is remarkable since known catalysts based on Ni NPs are less selective. Finally, we anticipate that the decoration of the surface of BP flakes with NPs may induce peculiar electronic and adsorption properties in comparison to pristine BP, which could be exploited for the fabrication of new electronic devices as for instance next-generation gas sensors.

This work was financially supported by EC through the project PHOSFUN "Phosphorene functionalization: a new platform for advanced *multifunctional materials*" (ERC ADVANCED GRANT N. 670173 to M.P.). F.T. thanks CNR-NANO for funding the SEED project SURPHOS. S.H. acknowledges support from Scuola Normale Superiore, project SNS16_B_HEUN – 004155. Thanks are expressed also to Italian Ministry of Education and Research (MIUR) under project Beyond-Nano (PON a3_00363).

## Notes and references

# Decoration of exfoliated black phosphorus with nickel nanoparticles and application in catalysis

Maria Caporali,[a,]* Manuel Serrano-Ruiz,[a] Francesca Telesio,[b] Stefan Heun,[b] Giuseppe Nicotra,[c] Corrado Spinella,[c] Maurizio Peruzzini[a,]*

[a] CNR-ICCOM, Via Madonna del Piano 10, 50019 Sesto Fiorentino, Italy.
[b] NEST, Istituto Nanoscienze-CNR and Scuola Normale Superiore, Piazza S. Silvestro 12, 56127 Pisa, Italy.
[c] CNR-IMM, Istituto per la Microelettronica e Microsistemi, Strada VIII, 5, 95121 Catania, Italy.

*corresponding authors, e-mail: maria.caporali@iccom.cnr.it; mperuzzini@iccom.cnr.it

**Experimental Section**

**Materials**

All reactions were performed under nitrogen atmosphere using standard air-free techniques. Anhydrous dimethylsulfoxide, anhydrous dimethylformamide, hexane, trioctylphosphine (97%), Ni(acac)$_2$ (95%), oleylamine (70%) were used as received from Sigma Aldrich. Black phosphorus[1] and nickel nanoparticles[2] were prepared according to literature procedures.

**Preparation of few-layer black phosphorus (2D BP)**: Black phosphorus (BP) (5.0 mg) was put in a vial and dry dimethylsulfoxide (5.0 mL) and deoxygenated distilled water (3 μL - 5 μL) were added. The vial was sealed under nitrogen and exposed to ultrasound for 120 hours at a power of 200 W and 37 kHz. The sonication bath was constantly kept at 37 °C and in the dark. Afterwards, the vial was opened, degassed acetone was added (25.0 mL), and the suspension was centrifuged for 90 minutes at 6000 rpm. The colorless surnatant was removed, fresh acetone (25.0 mL) was added, and again the suspension was centrifuged at 6000 rpm for 90 min. The resulting dark grey powder was re-suspended by sonication (approximately one minute) in the desired organic solvent, (DMSO, THF, acetone, 2-propanol) and stored under nitrogen in the dark. The purity of the material was checked by ICP-MS.

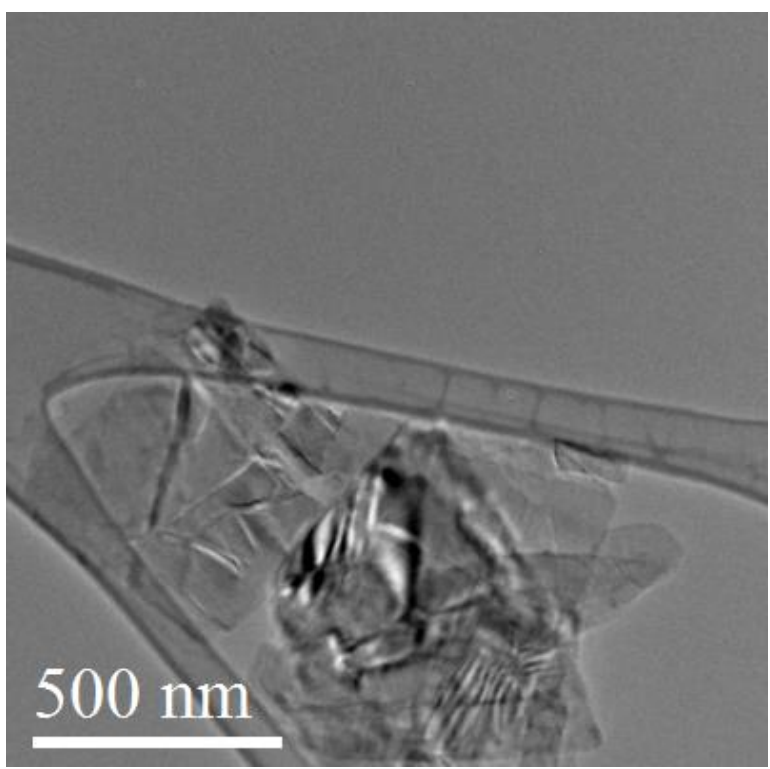

**Figure S1**. Low-magnification image of BP flakes, anchored after dripping on Lacey carbon support film, imaged in TEM-BF mode.

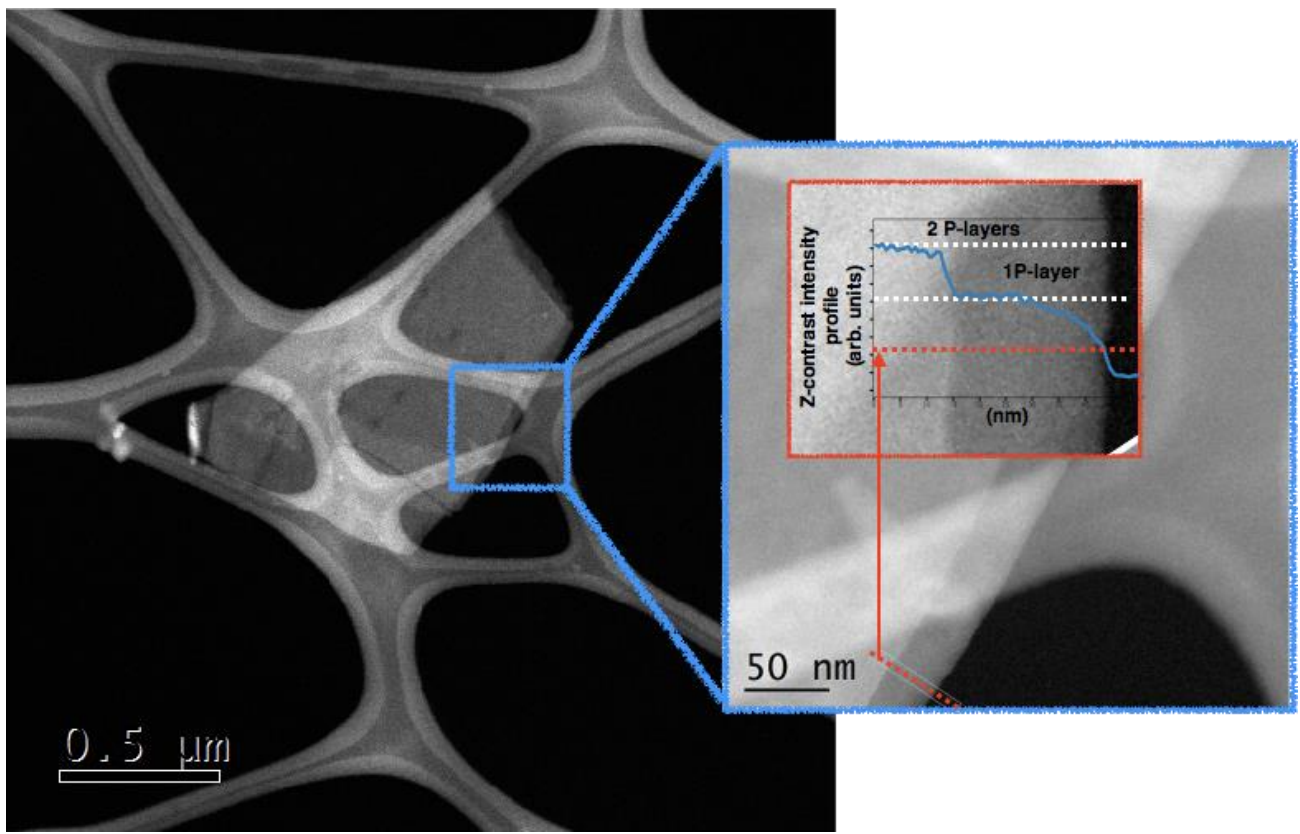

**Figure S2**. Low-magnification STEM Z-contrast image of a stack of two BP flakes on Lacey carbon support film. (inset) Line intensity scan on higher magnified image showing the presence of only two BP flakes, as underlined by the two steps, the first step between the two flakes, and the second one between the lower flake and the vacuum.

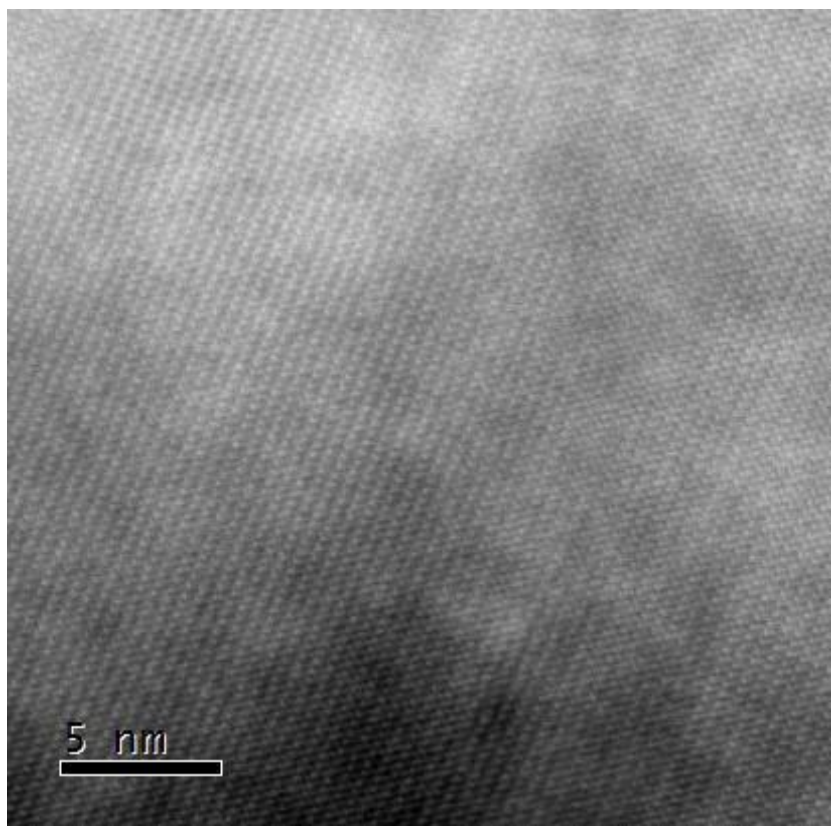

**Figure S3**. STEM Z-contrast image at atomic resolution showing the crystalline structure of a BP flake

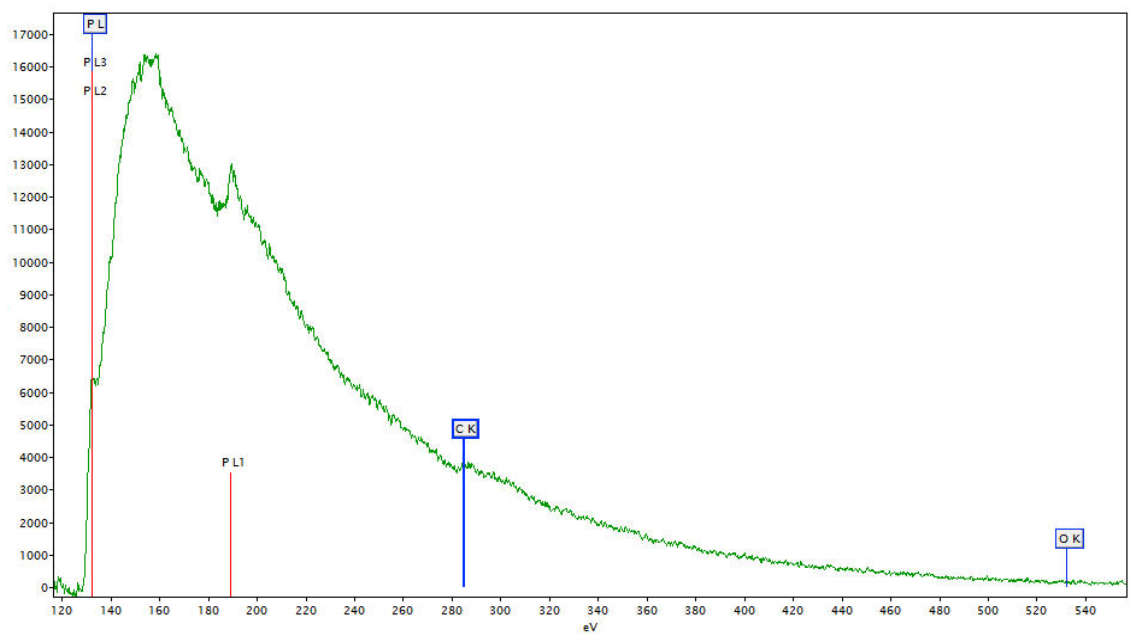

**Figure S4**. EELS spectrum of exfoliated black phosphorus flake. No oxygen is detected, indicating that the flake is not oxidized. The weak C K-edge peak arises from carbon supporting TEM grid.

**Synthesis of Ni NPs.**

To solid Ni(acac)$_2$ (50.0 mg, 0.194 mmol), oleylamine (640 µL, 1.362 mmol) and trioctylphosphine (70 µL, 0.152 mmol) were added under nitrogen. The mixture was heated up to 220°C in 15 minutes under nitrogen and kept at this temperature for two hours, following a published procedure.[2] The average Ni NPs diameter was 11.4 nm with a standard deviation of 1 nm, see figure S5 below.

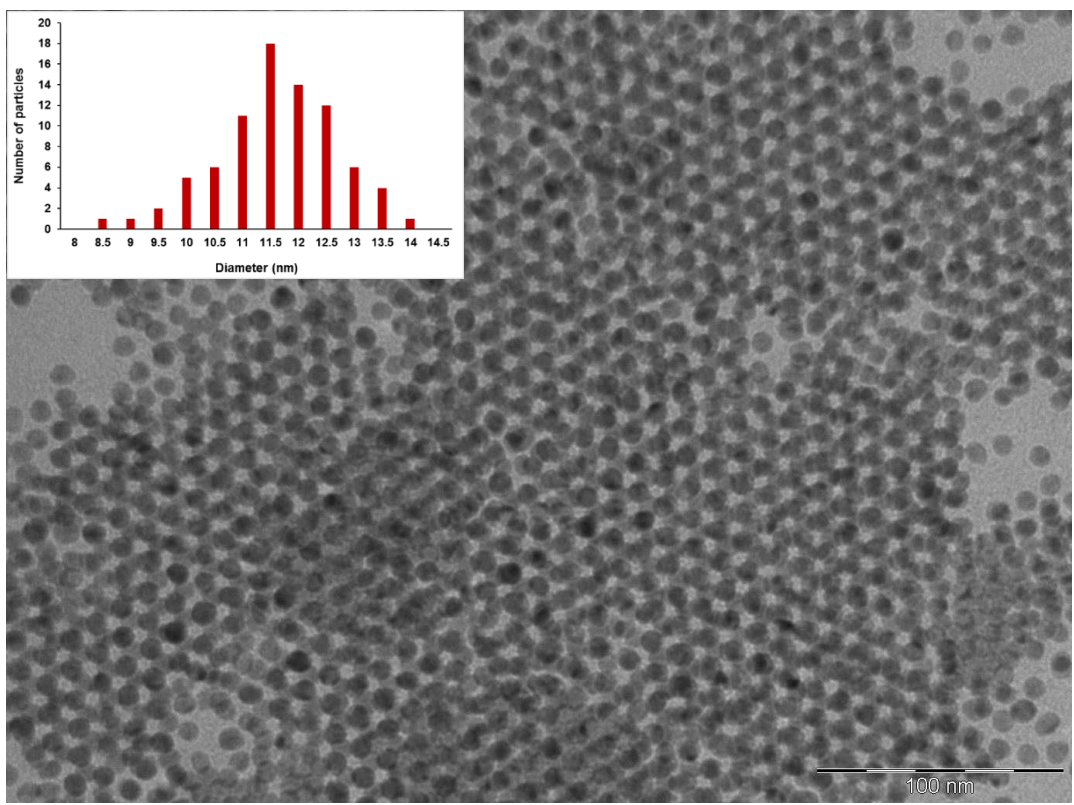

**Figure S5.** Bright field TEM image of Ni NPs and relative size distribution histogram. Scale bar: 100 nm.

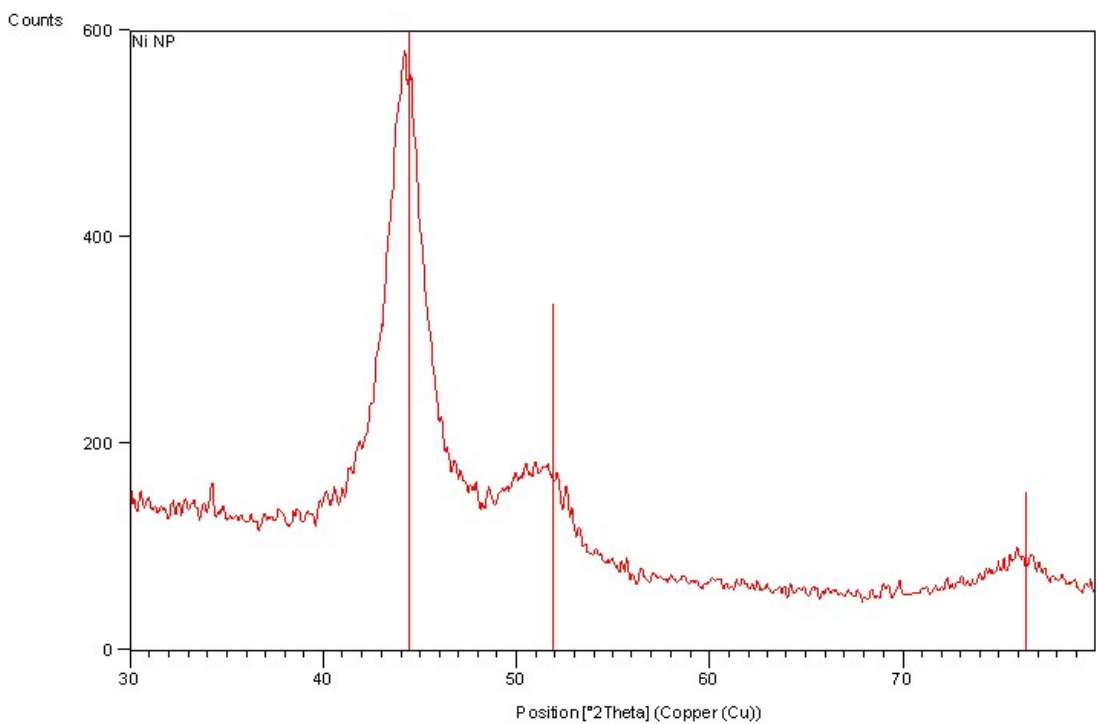

**Figure S6.** Powder X-Ray diffraction of Ni NPs.

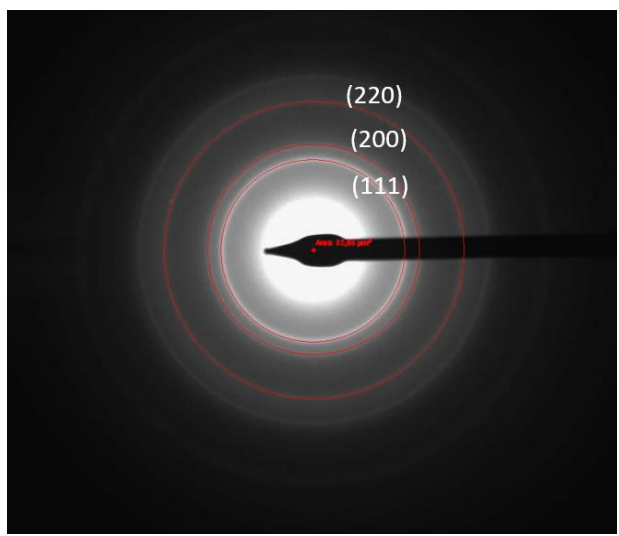

**Figure S7.** SAED (selected area electron diffraction) pattern of Ni NPs. The diffraction rings correspond to face-centered cubic nickel.

**Immobilization of Ni NPs on phosphorene (Ni/2D BP)**

All the samples used in the catalysis and shown in the pictures were prepared as follows:

To a freshly prepared suspension of few-layer BP in THF (1.2 mL, 1.2 mg, 0.0387 mmol "P") a black colloidal solution of nickel nanoparticles dispersed in toluene (0.29 mL, 0.018 M, 0.00531 mmol "Ni", P/Ni = 7.3) was added drop-wise under nitrogen at room temperature. After stirring for 30 minutes, degassed acetone (10.0 mL) was added, and the mixture was centrifuged at 8000 RPM for 20 minutes. The black residue was washed once more with acetone (10 mL), dried under a stream of nitrogen and re-suspended in THF. The Ni content of the isolated catalyst was measured by ICP-MS and resulted Ni to be 20.6 wt.%.

The same procedure was followed for the preparation of the catalyst having molar ratio P/Ni = 10.5, with a Ni content of 15.2% in weight as measured by ICP. The molar concentration of the colloidal solution of nickel nanoparticles was also assessed by ICP.

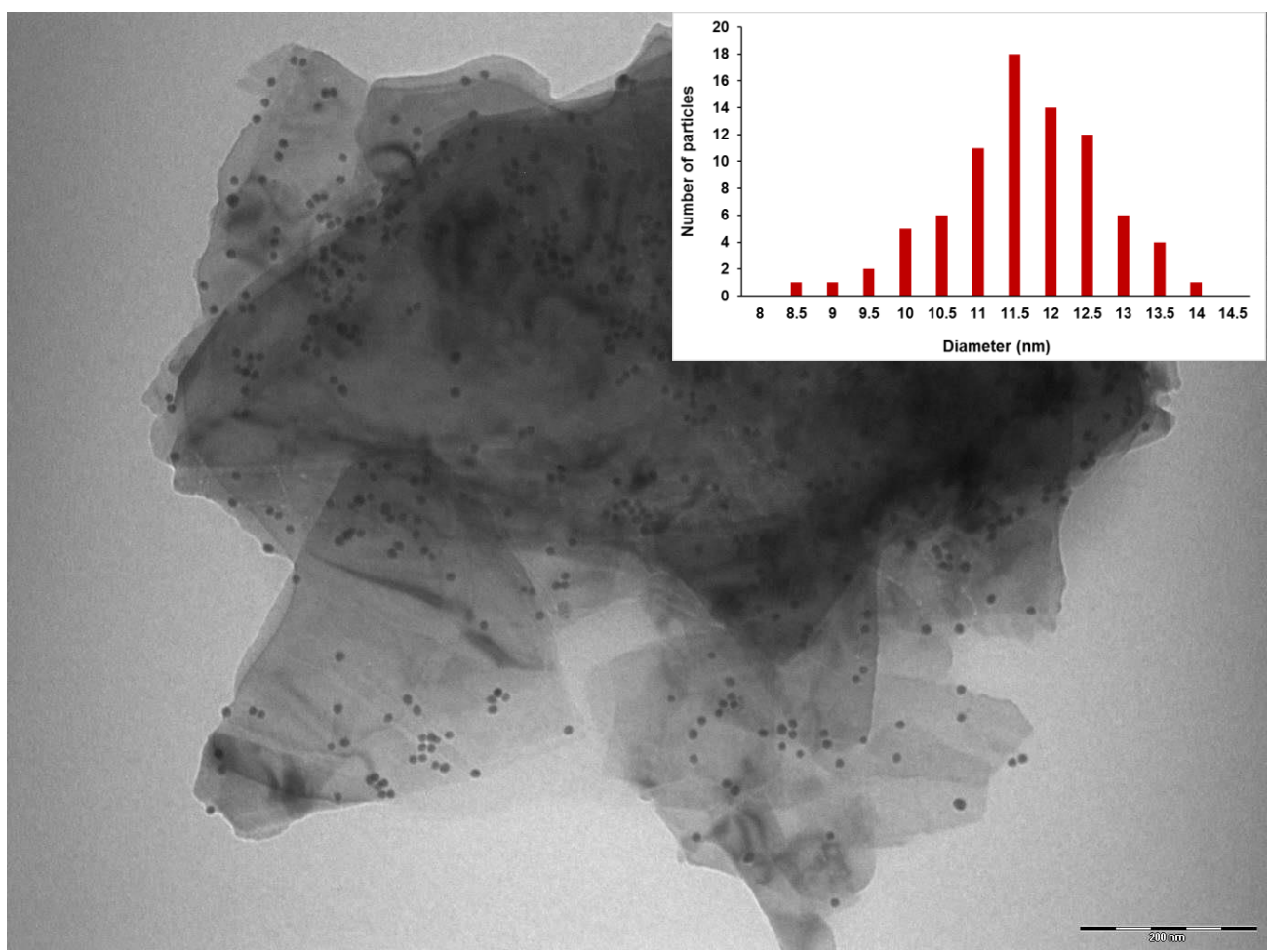

**Figure S8**. Bright field TEM image of Ni/2D BP, P: Ni molar ratio equal to 7.3:1 and relative size distribution histogram. Scale bar: 200 nm.

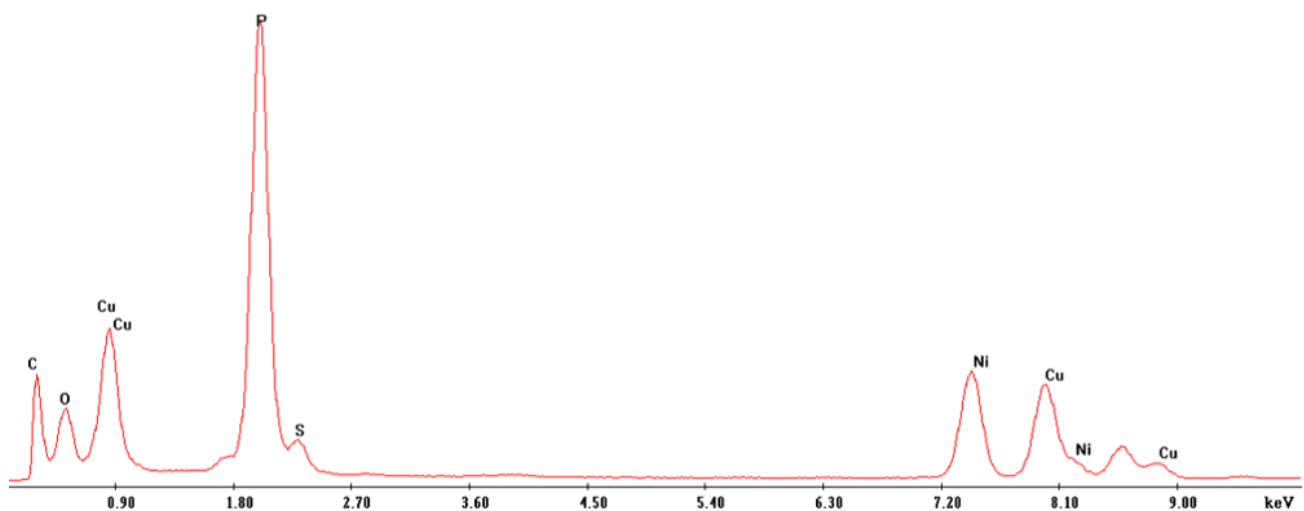

**Figure S9**. Energy dispersive X-ray spectrum of Ni/2D BP. The small peaks of C, O, and S can be attributed to a residual amount of DMSO. The Cu peaks originate from the Cu TEM grid on which the sample was placed.

a)
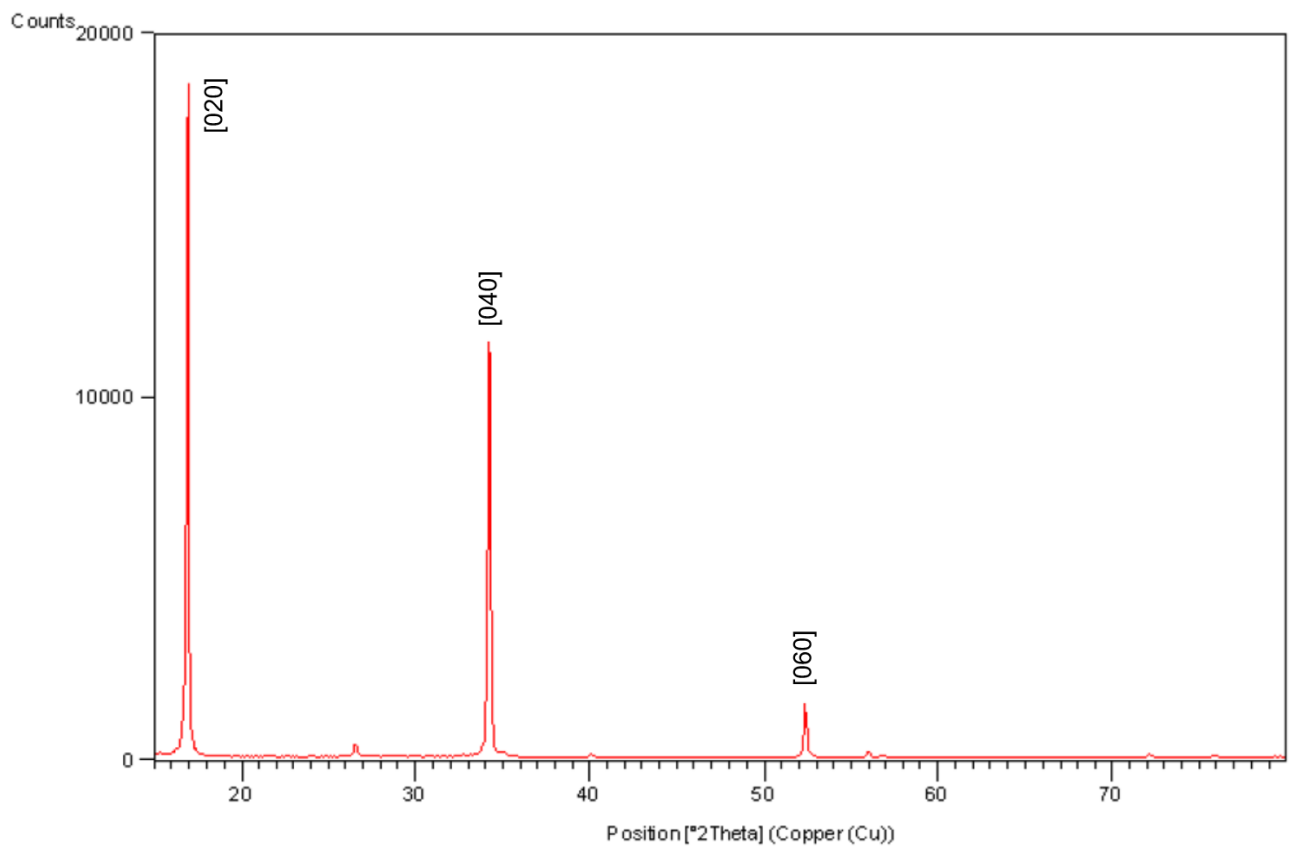

b)

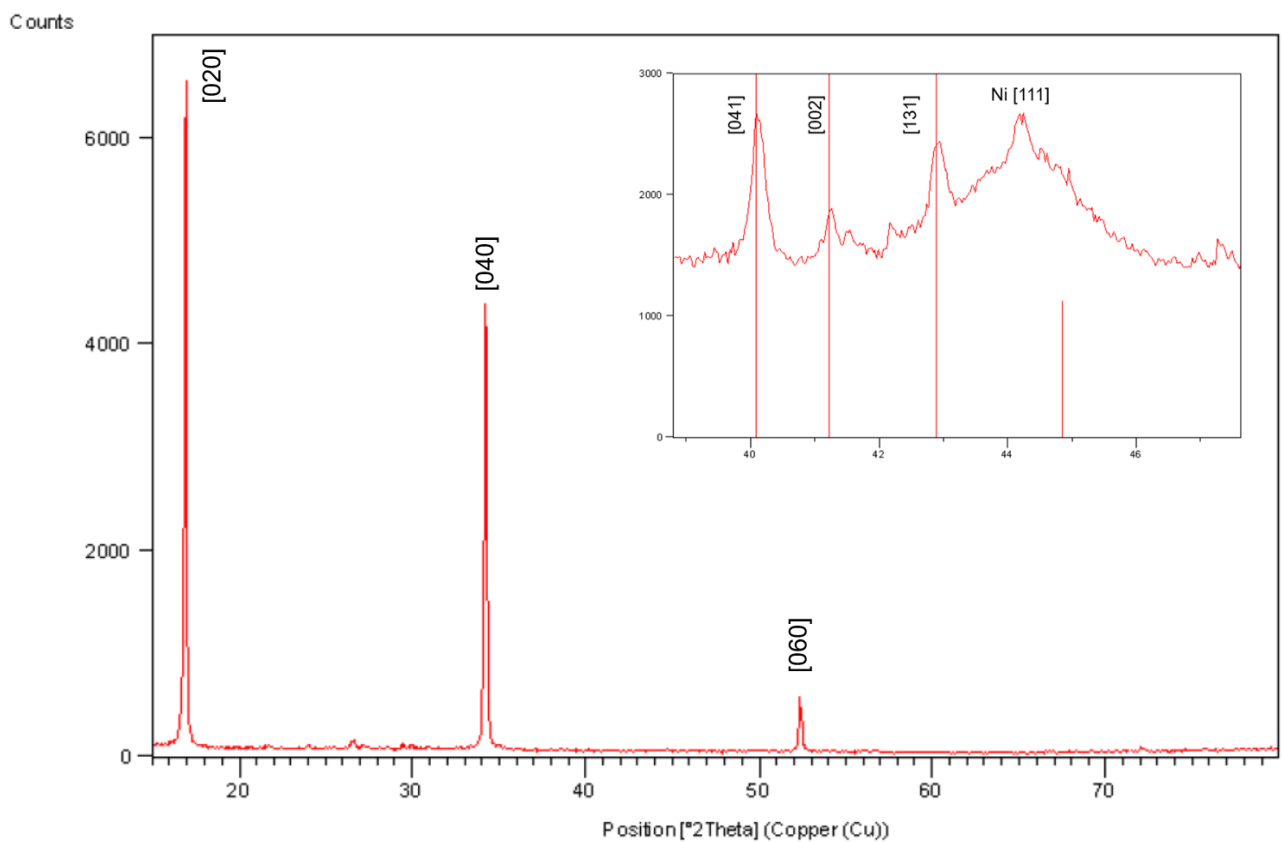

**Figure S10.** a) Powder X-ray diffraction of exfoliated black phosphorus. b) Powder X-ray diffraction of Ni/2D BP. Inset of the region from $2\theta = 39°$ to $47°$, obtained after a prolonged acquisition. Peaks marked with a bar are characteristic of black P, the broad peak at $2\theta = 44.4°$ is attributed to nickel.

**Catalytic hydrogenations.** In a typical experiment, a 50 mL stainless steel Parr reactor was equipped with a vial, containing a magnetic stirrer, and charged with a suspension of Ni NPs supported on 2D BP in tetrahydrofuran/toluene (3:1) under an inert atmosphere. Phenylacetylene was added, then the autoclave was sealed, purged with hydrogen (3 times) and then pressurized with hydrogen up to 10 bar. The autoclave was kept stirring at 80 °C in an oil bath preheated at 80 °C for the desired time. After that, the autoclave was cooled down to room temperature with ice and depressurized. Ultracentrifugation of the reaction mixture for 5 minutes at 10000 RPM allowed the heterogeneous catalyst to settle down. The colourless supernatant liquid was analyzed by gas chromatography and mass spectrometry. To the black solid catalyst, a fresh solution of phenylacetylene in tetrahydrofuran/toluene (3:1) was added, and a new catalytic run was launched.

**Table S1** Comparison of catalytic activity in the semihydrogenation of phenylacetylene.

| Entry | Conversion (%) | Selectivity to styrene (%) |
|---|---|---|
| Ni NPs | 100.0 | 78.6 |
| 2D BP | 0.0 | 0.0 |
| Ni /2D BP | 93.2 | 92.8 |

**Reaction conditions**: 10 bar $H_2$, 80 °C, 1 hour, Ni NPs = 1.8 mol%, phenylacetylene: 0.30 mmol, 1.51 mg Ni/2D BP (20.6 wt% Ni). Both conversion and selectivity were evaluated by gas chromatography and gas chromatography interfaced with mass spectrometry (GC-MS).

**Characterization**

**Transmission electron microscopy.** TEM studies were carried out using a Philips instrument operating at an accelerating voltage of 100 kV. Few drops of the Ni NPs on 2D BP suspension in isopropanol were placed on the TEM lacey copper/carbon grid, air dried, and measured.

Atomic-resolution characterization by STEM was performed with a probe spherical aberration-corrected JEOL ARM200CF, equipped with a Ceos hexapole-type Cs corrector, named CESCOR. The electron gun is a cold-field emission gun with an energy spread of 0.3 eV. The microscope was operated at a primary beam energy of 60 keV at which it is capable to deliver a probe size of 1.1 Å. Images were acquired in Z-contrast mode, by using the Gatan HAADF detector with a beam illumination angle of 33 mrad and a collection angle of 90 mrad. To improve the signal-to-noise ratio of the HAADF-STEM, a low-pass filtering of the images was applied. A last generation post-column energy filter, GIF Quantum ER, was used as EELS spectrometer, and EELS chemical maps were nearly simultaneously acquired to the HAADF-STEM images by using the 2D Spectrum Imaging (SI) capability.[3]

**Gas Chromatography** analyses were performed on a Shimadzu GC-14A gas chromatograph (with apolar column) equipped with flame ionization detector and a SPB-1 Supelco fused silica capillary column (30 m, 0.25 mm i.d., 0.25 μm film thickness).

**Powder X-ray diffraction** data were collected with an X'Pert PRO diffractometer with Cu-Kα radiation (λ = 1.5418).

**Raman** spectroscopy was performed using a Renishaw inVia system equipped with a 532 nm laser and a motorised stage for 2D mapping of samples. A laser spot size of approximately 2 μm in diameter was used.

**UV-Vis absorption spectra** were recorded using a Shimadzu 2600 spectrophotometer. Samples were tested using tetrahydrofuran as solvent in stoppered quartz cuvettes having 10 mm path length.

**Inductively coupled plasma mass spectrometry** (ICP-MS) measurements were performed with an Agilent 7700 Series spectrometer. Samples followed a microwave-assisted digestion in Nitric acid for trace analysis. Then, different dilutions of each sample with water for trace analysis were prepared, in order to obtain concentrations in the sensitivity range of the instrument for the elements under investigation (namely Ni and P). Standards at different concentrations have also been prepared and measured contextually to sample measurements, in order to obtain a calibration curve for each element under investigation.

**Atomic force microscopy** (AFM) measurements were performed with a Bruker Dimension Icon Atomic Force Microscope, in pick force mode. Samples for AFM were prepared by drop cast on a Si/SiO$_2$ substrate. The DMSO drop was left in contact with the substrate for 90 minutes, then the sample was washed with acetone and dried in vacuum overnight. Samples prepared with this method have almost no solvent traces, at least far from the edges, and quite small aggregates, as shown in Figure S11 (a). Nevertheless, re-aggregation of functionalized BP flakes could not be completely prevented during the deposition procedure. During AFM, we approached in regions where no aggregates were visible in the optical microscope, and found many small structures, shown in Figure S11 (b), which are likely to be single BP flakes or few flake-aggregates. These structures have a typical size of a few hundred nanometers and a thickness of up to 50nm, with a substrate coverage of 2-4%. An image of a typical flake is shown in the main text, in Figure 6.

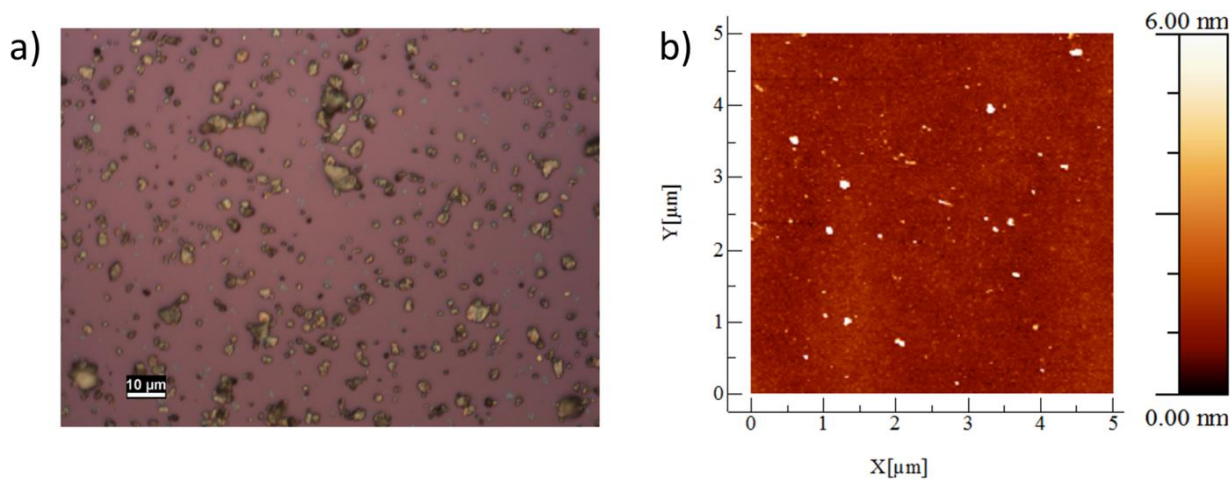

**Figure S11**. (a) Optical microscopy image of a sample prepared for AFM by drop casting from a Ni/2D BP solution and (b) an AFM image (size 5μm x 5μm) of the same sample.

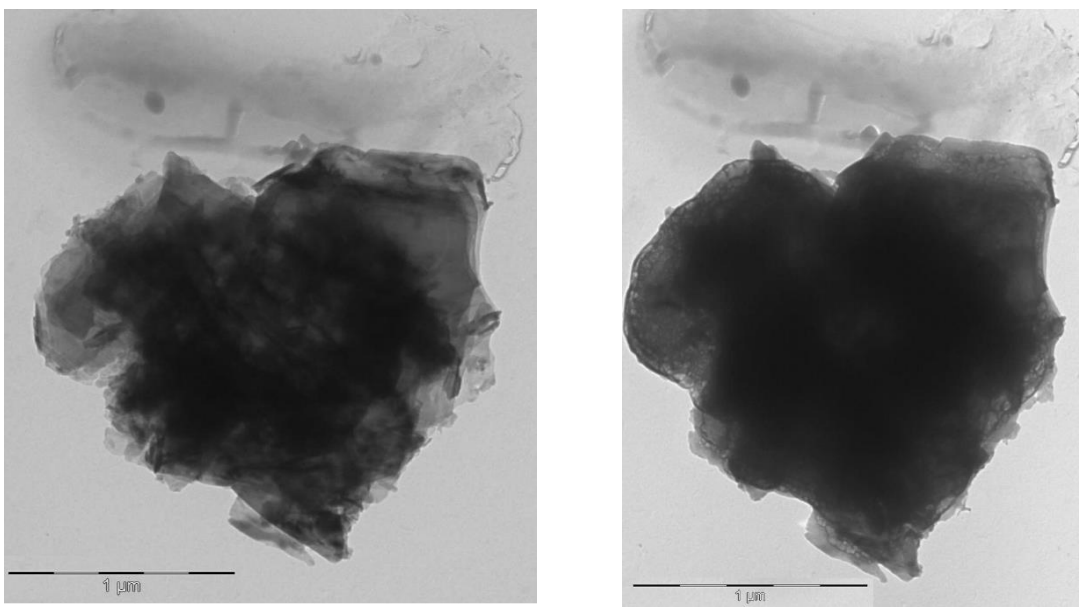

**Figure S12.** Bright field TEM images of pristine 2D BP. Left: freshly deposited BP flake on a TEM grid; right: the same flake that looks heavily degraded after one week exposure to ambient conditions in the dark. Scale bar: 1 μm.

a)

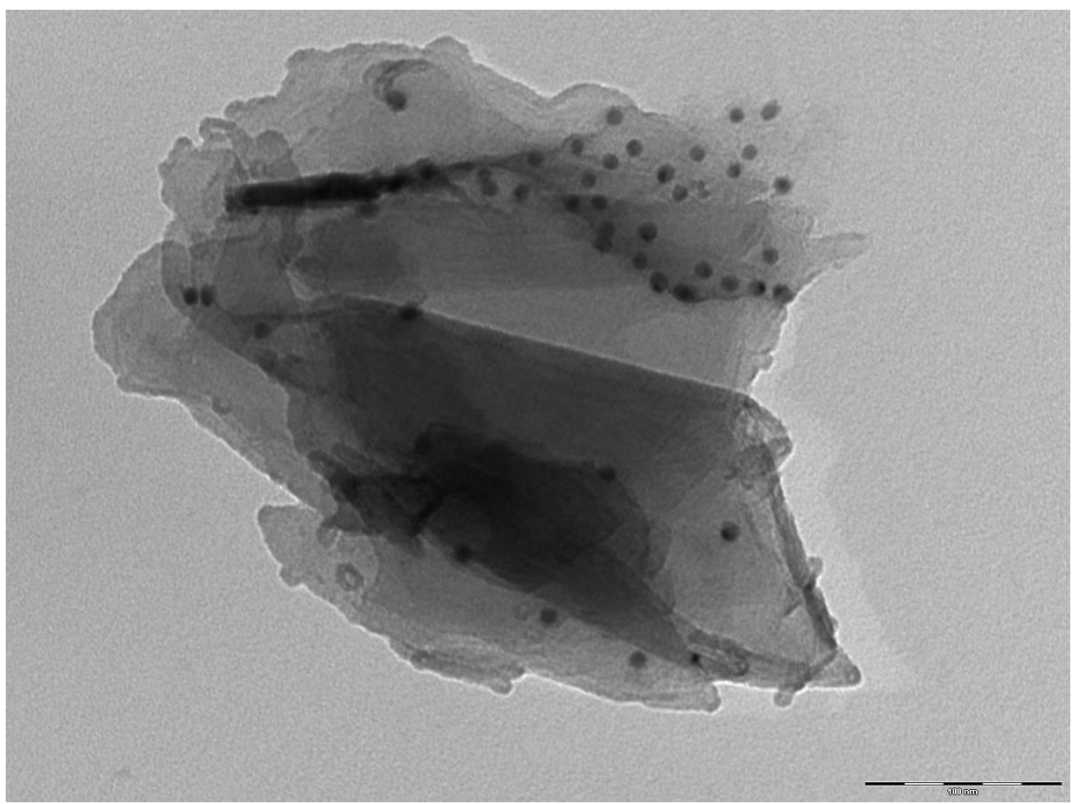

b)

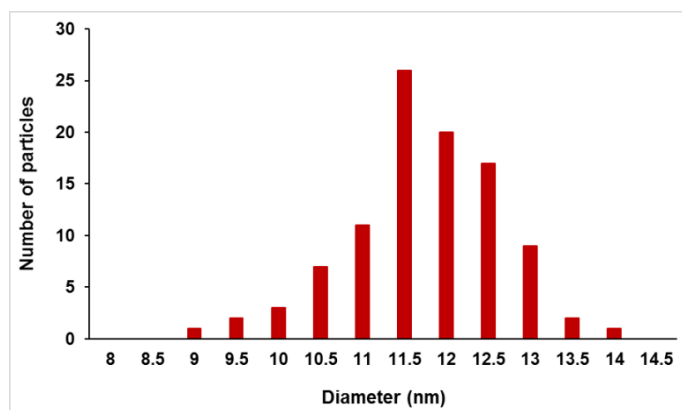

**Figure S13.** a) Bright field TEM image of Ni/2D BP after catalytic tests, scale bar: 100 nm; and b) relative size distribution histogram.

---

[1] T. Nilges, M. Kersting, T. Pfeifer, *J. Solid State Chem*. 2008, **181**, 1707-1711.

[2] S. Carenco, C. Boissière, L. Nicole, C. Sanchez, P. Le Floch, N. Mézailles, *Chem. Mater*. 2010, **22**, 1340-1349.

[3] G. Nicotra, Q. M. Ramasse, I. Deretzis, A. La Magna, C. Spinella, F. Giannazzo *ACS Nano*, 2013, **7**, 3045-3052.